\documentclass[conference]{IEEEtran}

\usepackage{cite}
\usepackage{amsmath,amssymb,amsfonts}
\usepackage{algorithm}
\usepackage{algpseudocode}%
\usepackage{graphicx}
\usepackage{hyperref}
\usepackage{textcomp}
\usepackage{xcolor}
\usepackage{subcaption}
\usepackage{multirow}
\usepackage{siunitx}
\usepackage{lipsum}
\usepackage{mathtools}
\usepackage{cuted}
\usepackage{stfloats}
\usepackage{bbm} 
\usepackage{balance}

\usepackage{enumitem}

\usepackage{amsthm}

\usepackage{tabularx}
\makeatletter
\newcommand{\multiline}[1]{%
  \begin{tabularx}{\dimexpr\linewidth-\ALG@thistlm}[t]{@{}X@{}}
    #1
  \end{tabularx}
}
\makeatother

\usepackage[font=small]{caption}

\allowdisplaybreaks
\IEEEoverridecommandlockouts

\begin{document}

\title{ 
Learning-Based Adaptive User Selection in Millimeter Wave Hybrid Beamforming Systems
}

\author{
\IEEEauthorblockN{Junghoon~Kim and Matthew~Andrews}
\IEEEauthorblockA{Nokia Bell Labs, Murray Hill, NJ, USA}
\thanks{Work performed while J.~Kim was a summer intern from Purdue University. Emails: kim3220@purdue.edu, matthew.andrews@nokia-bell-labs.com.
The authors thank Fran\c{c}ois Durand for providing our experimental platform.
}
}


\maketitle

\begin{abstract}

%
%
We consider a multi-user hybrid beamforming system, where the multiplexing gain is limited by the small number of RF chains employed at the base station (BS). To allow greater freedom for maximizing the multiplexing gain, it is better if the BS selects and serves some of the users at each scheduling instant, rather than serving all the users all the time.

We adopt a two-timescale protocol that takes into account the mmWave characteristics, where at the long timescale an analog beam is chosen for each user, and at the short timescale users are selected for transmission based on the chosen analog beams. The goal of the user selection is to maximize the traditional Proportional Fair (PF) metric. 
However, this maximization is non-trivial due to interference between the analog beams for selected users. We first define
a greedy algorithm and a ``top-$k$'' algorithm,
and then propose a \textit{machine learning} (ML)-based user selection algorithm to provide an efficient trade-off between the PF performance and the computation time.
Throughout simulations, we analyze the performance of the ML-based algorithms under various metrics,
and show that it gives an efficient trade-off in performance as compared to counterparts. 


\end{abstract}


\section{Introduction}

Millimeter wave (mmWave) frequency bands 
hold considerable promise for 
5G-and-beyond wireless communications~\cite{hur2013millimeter, rusek2012scaling}.
To combat the high path-loss in mmWave, a base station (BS) typically  employs a large antenna array and conducts beamforming to improve the communication efficiency. 
Initial investigations considered analog beamforming, 
where the beams are formed by phase-shifters at each antenna element, 
and only a single radio frequency (RF) chain is needed~\cite{hur2013millimeter}. However, analog beamforming in isolation is limited in how well it can
manage the multi-user interference.
On the other hand, 
fully-digital beamforming requires the BS to employ an individual RF chain for each antenna element, so that the transmit/receive signals are processed in the digital domain~\cite{rusek2012scaling}.
Employing many RF chains in large antenna array mmWave systems is not realistic since it drastically increases implementation costs and power consumption. 

To benefit from both analog and digital beamforming,
a \textit{hybrid} beamforming structure has been proposed for mmWave systems, where the BS conducts both the analog and digital beamforming with a number of RF chains much less than the number of antenna elements~\cite{alkhateeb2015limited}.
%
%
However, the small number of RF chains limits the multiplexing gain of the multi-user hybrid beamforming systems~\cite{alkhateeb2015limited}.
This implies that a few RF
chains limit the maximum number of users to be served by the BS at a time.
Due to this hardware constraint in hybrid beamforming, the BS  conducts \textit{user selection} (or scheduling) to choose some of the users to serve at each time instant. 

Recently, the user selection problem has been studied within the framework of multi-user hybrid beamforming systems.
The work~\cite{vu2022beam} addresses joint user selection and hybrid beamforming to maximize the sum-rate over the selected users given a perfect channel state information (CSI). However, the perfect CSI assumption is not realistic in mmWave systems.
In \cite{kwon2016joint,kim2017hybrid}, the authors consider a more practical CSI acquisition by employing an analog beamforming codebook at the BS, and aim to maximize the sum-rate through user selection and hybrid beamforming.
However, the aforementioned works only maximize the sum-rate in a {\em single} time frame and do not consider long-term user fairness. 
Furthermore, the nature of the time-varying channels has not been considered in the previous works, which necessitates the BS to conduct an \textit{adaptive} user selection
according to the channel variations and the updated service priorities for serving the users fairly.

In this work, we address adaptive user selection for mmWave hybrid beamforming systems.
We first tailor the signal model for the hybrid beamforming structure by incorporating the varying number of users.
We then formulate the proportional fairness (PF) maximization problem 
for
the mmWave hybrid beamforming system, where the PF metric balances between the user fairness and the data rates of the users~\cite{andrews2007survey,stolyar2005maximizing}.
We adopt a two-timescale protocol that takes into account the mmWave characteristics~\cite{liu2019novel,cai2020two}, where at the long timescale an analog beam is chosen for each user, and at the short timescale users are
selected for transmission and digital beamforming is conducted, both based on the chosen analog beams. 
Hence our setup differs from the standard PF scheduling problem since the user channel conditions are not defined solely by exogenous propagation conditions, but also by the analog beams that we choose. 





Solving the user selection problem in the hybrid system faces the following three challenges: (i) there are a huge number of combinations of a potential user set, (ii) the user selection should be conducted fast enough within
the short timescale, i.e., a few ${\rm msecs}$, which is the channel coherence time in mmWave systems~\cite{akdeniz2014millimeter,va2016beam},
and (iii) the user selection should entail the joint consideration of the user channel conditions, the chosen analog beams, the updated service priorities, and the digital beamforming.
For user selection,
our first baseline is a greedy algorithm that iteratively adds users according to the PF metric.
However, it is computationally expensive, which may make it non-realistic to operate within the short timescale,
although it yields a good PF.
Another baseline is a simple ``top-$k$'' heuristic, which provides a short operation time while yielding a low PF.
To provide an efficient trade-off between the PF performance and the computation time,
we propose to exploit machine learning (ML) for user selection to \textit{learn} the relationship between the available information (i.e., the measured user channels,  chosen analog beams, and  updated service priorities) and the best user set.
Through simulations, we show that the ML-based algorithm gives an efficient trade-off in performance as compared to the alternatives.

\section{System Model for Adaptive User Selection}

\begin{figure}[t]
  \includegraphics[width=\linewidth]{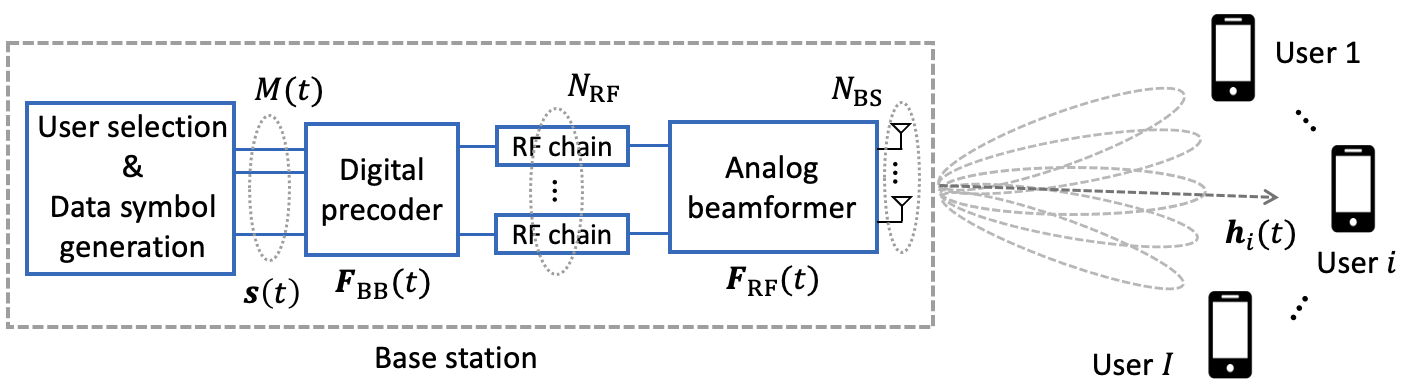}
  \vspace{-2mm}
  \centering
  \caption{A multi-user mmWave hybrid system in downlink, where a BS employs a hybrid analog/digital beamforming structure and communicates with  $I$ users. At each time, the BS selects some of the users adaptively and conducts the hybrid beamforming accordingly.
  }
  \label{fig:system}
\end{figure}

We consider a downlink multi-user system, where a BS serves a total of $I$ users in a network, as shown in Fig.~\ref{fig:system}.
We define the set of user indices as $\{1,2,...,I \}$.
We adopt a hybrid beamforming architecture at the BS, where the number of BS antennas and RF chains is $N_{\rm BS}$ and $N_{\rm RF}$, respectively, and $N_{\rm BS} > N_{\rm RF}$, while each user has a single antenna.
We assume that the BS communicates with a user via \textit{only one data stream} 
as in~\cite{alkhateeb2015limited} and we let $N_{\max}$ be the maximum number of users that can be served simultaneously by the BS.
It is known that the spatial multiplexing gain of the multi-user hybrid
beamforming system is limited by $\min(N_{\rm RF}, I)$~\cite{alkhateeb2015limited}.
Considering a few RF chains as $N_{\rm RF} < I$,
it is reasonable to set 
$N_{\max} = N_{\rm RF}$.
%
We consider a block-fading channel model where the channels are fixed in each channel block. 
We denote the channel block index as $t$ where $t \in \{1, 2, ...\}$.

At each time $t$, the BS serves $M(t)$ users among a total of $I$ users, where $M(t) \le N_{\max}$.
We define the set of indices of the selected users at time $t$ as 
\begin{equation}
\mathcal{M}(t) = \{ i_{1}(t), ..., i_{M(t)}(t) \} \subset \{1,...,I\},
\end{equation}
where $\vert \mathcal{M}(t) \vert = M(t)$ and $1 \leq i_{1}(t) < i_{2}(t) < ... < i_{M(t)}(t) \leq I$.
%
For the selected $M(t)$ users,
the BS transmits the symbols 
\begin{equation}
{\bf s}(t)= [{s}_{i_1(t)}(t), ..., {s}_{i_{M(t)}(t)}(t)]^T  \in \mathbb{C}^{M(t) \times 1},
\end{equation}
where  ${s}_{i}(t) \in \mathbb{C}$ is intended for user $i$, $i \in \mathcal{M}(t)$.\footnote{For example, if $\mathcal{M}(t)=\{2,4,5\}$, we have $M(t)=3$, $i_1(t)=2$, $i_2(t)=4$, and $i_3(t)=5$. Correspondingly, we have
${\bf s}(t)= [s_2, s_4, s_5]^T$.}
We assume  ${\mathbb E}[\vert s_i(t) \vert^2]=1$ and ${\mathbb E}[ s_{i}^*(t) s_{j}(t)]=0$ for $i \neq j$, where $i,j \in  \mathcal{M}(t)$. 
We consider the digital (baseband) precoder at time $t$ as 
\vspace{-1mm}
\begin{equation}
    \hspace{-1mm} {\bf F}_{\rm BB}(t) = [{\bf f}_{{\rm BB},i_1(t)}(t), ..., {\bf f}_{{\rm BB},i_{M(t)}(t)}(t)] \in \mathbb{C}^{M(t) \times M(t)},
    \label{eq:BBprecoder}
    \vspace{-1mm}
\end{equation}
followed by the analog (RF) precoder
\begin{equation}
    {\bf F}_{\rm RF}(t) = [{\bf f}_{{\rm RF},i_1(t)}(t), ..., {\bf f}_{{\rm RF},i_{M(t)}(t)}(t)] \in \mathbb{C}^{ N_{\rm BS} \times M(t)}.
    \label{eq:RFprecoder}
\end{equation}
%
At time $t$, the BS transmits the signal  ${\bf x}(t) \in \mathbb{C}^{N_{\rm BS} \times 1}$ as
\begin{align}
    {\bf x}(t) & = {\bf F}_{\rm RF}(t) {\bf F}_{\rm BB}(t) {\bf s}(t)
    \nonumber
    \\
    & = {\bf F}_{\rm RF}(t) \sum\limits_{j \in \mathcal{M}(t)}{\bf f}_{{\rm BB},j}(t) {s}_j(t).
\end{align}
%
The BS has the transmit power constraint ${\mathbb E}[ \vert {\bf x}(t) \vert^2 ] \le P$, where $P \in \mathbb{R}_+$ is the power limit.
We note that the sizes of  ${\bf F}_{\rm RF}(t)$, ${\bf F}_{\rm BB}(t)$, and ${\bf s}(t)$ vary depending upon the number of selected users at time $t$, i.e., $M(t)$.\footnote{Since we assume that each user uses only a single data stream, the number of \textit{active} RF chains will be $M(t) \le N_{\rm RF}$. In other words, $N_{\rm RF} - M(t)$ RF chains remain unused at time $t$.
For future works, it would be interesting to use multiple data streams for each user and fully utilize the RF chains in order to further improve the performance.}

We denote the channel from the BS to user $i$ at time $t$ as ${\bf h}_i(t) \in \mathbb{C}^{N_{\rm BS} \times 1}$.
User $i$, $i \in \mathcal{M}(t)$, receives the signal 
${ z}_i(t) \in \mathbb{C}$
at time $t$, given by
\begin{align}
    {z}_i(t) &= {\bf h}_i^H(t) {\bf x}(t) + {n}_i(t),
    \nonumber
    \\
    &= {\bf h}_i^H(t) {\bf F}_{\rm RF}(t) \sum\limits_{j \in \mathcal{M}(t)} {\bf f}_{{\rm BB},j}(t) {s}_j(t)     + {n}_i(t),
    \label{eq:signal_recevied}
\end{align}
where ${n}_i(t) \in \mathbb{C}$ is the 
Gaussian noise 
following ${n}_i(t) \sim \mathcal{CN}({0}, \sigma_i^2)$ with noise variance $\sigma_i^2$.

%
%
From \eqref{eq:signal_recevied},
we can define the signal to interference plus noise ratio (SINR) for user $i$, $i \in \mathcal{M}(t)$, at time $t$ as
\begin{align}
    {\rm SINR}_i(t) =  \frac{  \big\vert  {\bf h}_i^H(t) {\bf F}_{\rm RF}(t)  {\bf f}_{{\rm BB},i}(t)  \big\vert^2 }
    {\sum\limits_{j \in \mathcal{M}(t)\setminus \{i\}}  \big\vert  {\bf h}_i^H(t) {\bf F}_{\rm RF}(t)  {\bf f}_{{\rm BB},j}(t)  \big\vert^2 +   \sigma_i^2}.
    \label{eq:SINR}
\end{align}
The SINR for the non-selected user $i \notin \mathcal{M}(t)$ is set to zero.
We can then define the data rate of any user $i$ as
\begin{align}
    r_i(t) = \begin{cases}
    \log_2 \big(
    1 + {\rm SINR}_i(t)
    \big), & i \in \mathcal{M}(t), \\
    0, & i \notin \mathcal{M}(t),
    \end{cases}
    \label{eq:rate}
\end{align}
where we assume a unit bandwidth.

In mmWave, the BS typically adopts a \textit{codebook} for analog beamforming due to the RF hardware constraints~\cite{alkhateeb2015limited,kwon2016joint, kim2017hybrid}.
The codebook is defined as
$\mathcal{F} = \{{\bf g}_1, ..., {\bf g}_{\vert \mathcal{F} \vert}\}$ where ${\bf g}_k$ is the $k$-th analog beamforming vector in the set $\mathcal{F}$ and $\vert \mathcal{F} \vert$ is the cardinality of $\mathcal{F}$.
Once the user set $\mathcal{M}(t)$ is selected,
we design the corresponding analog beamforming vectors via selection from the codebook ${\mathcal F}$, i.e., ${\bf f}_{{\rm RF},i}(t) \in {\mathcal F}$, $i \in \mathcal{M}(t)$, which gives the analog precoder ${\bf F}_{\rm RF}(t)$ in \eqref{eq:RFprecoder}.
\section{Proportional Fairness (PF) Maximization}

We first formulate a long-term optimization problem to maximize
the PF in Sec.~\ref{ssec:PF:long}, and convert it to consecutive one-shot optimization problems in Sec.~\ref{ssec:PF:oneshot}.
Then, we discuss the challenges associated with solving it in Sec.~\ref{ssec:challenges}.


\vspace{-.5mm}
\subsection{Long Term PF Maximization}
\label{ssec:PF:long}
\vspace{-.6mm}

The \textit{cumulative} data rate $R_i(T)$ for user $i$ at time $T$ is 
defined by an exponential moving average of the data rates $\{r_i(t)\}_{t=1}^T$
as in~\cite{andrews2007survey}
\begin{equation}
    R_i(T) = (1-\delta) R_i(T-1) + \delta r_i(T),
    \label{eq:cum_rate}
\end{equation}
for some constant weight $\delta \in [0,1]$.
The PF metric is defined as $\sum_{i=1}^I \log R_i(T)$~\cite{andrews2007survey,stolyar2005maximizing}.
By maximizing the PF, any user will not be starved completely since $\log 0 = -\infty$.
Our goal is to maximize the PF
subject to the constraints imposed
by the hybrid beamforming system:
\begin{align}
    & \underset{\{{\mathcal{M}}(t)\}_{t=1}^{T}, \;
    \{{\bf F}_{\rm RF}(t)\}_{t=1}^{T}, \;
    \{{\bf F}_{\rm BB}(t)\}_{t=1}^{T}}{\rm maximize} \;\; \sum_{i=1}^I \log R_i(T)
    \label{opt:longterm:obj}
    \\
    & {\rm s.t.} \quad {\mathcal{M}}(t) \subset \{1,...,I \}, \quad \vert {\mathcal{M}}(t) \vert \le N_{\max}, \quad t=1,...,T,
    \nonumber
    \\
    & \hspace{.8cm} \sum_{j \in \mathcal{M}(t)}  \| {\bf F}_{\rm RF}(t) {\bf f}_{{\rm BB},j}(t) \|^2 \le P, \quad t=1,...,T,
    \nonumber
    \\
    & \hspace{.8cm}
    {\bf f}_{{\rm RF},j}(t) \in {\mathcal F}, \quad j \in \mathcal{M}(t),\; t=1,..,T,
    \label{opt:longterm:con}
\end{align}

\noindent
The first constraint in \eqref{opt:longterm:con} accounts for the user selection requirement. The second line in \eqref{opt:longterm:con} captures the constraint on the transmission power. The last constraint takes into account the codebook utilization for the analog RF beamforming.

The optimization problem in \eqref{opt:longterm:obj}-\eqref{opt:longterm:con} requires a solution over all future time periods, i.e., from $t=1$ to $T$, which is practically challenging to solve.
Fortunately, this long-term optimization problem can be decomposed into consecutive \textit{one-shot} optimization problems
at each time $t$~\cite{stolyar2005maximizing}, which will be discussed in the next subsection.

\subsection{One-Shot Optimization for PF Maximization}
\label{ssec:PF:oneshot}

It is known that maximizing the PF metric, $\sum_{i=1}^I \log R_i(T)$, is equivalent to maximizing the weighted sum of instantaneous data rates, $\sum_{i=1}^I r_i(t)/R_i(t-1)$, from $t=1$ to $T$~\cite{stolyar2005maximizing}. 
This one-shot metric provides a user fairness, 
in that
the BS would select the users that have not been served for a long time, i.e., the users with low $R_i(t-1)$. 
We then decompose
the long-term optimization problem  \eqref{opt:longterm:obj}-\eqref{opt:longterm:con} into $T$ consecutive
  one-shot optimization problems. The one-shot optimization problem at each time $t \in \{1,...,T\}$ is given by
\begin{align}
    & \underset{{\mathcal{M}}(t), \;
    {\bf F}_{\rm RF}(t), \;
    {\bf F}_{\rm BB}(t)}{\rm maximize} \quad \sum_{i=1}^I
    w_i(t) r_i(t)
    \label{opt:oneshot:obj}
    \\
    & {\rm s.t.} \quad 
    {\mathcal{M}}(t) \subset \{1,...,I \}, \quad \vert {\mathcal{M}}(t) \vert \le N_{\max}, 
    \nonumber
    \\
    & \hspace{.8cm} \sum_{j \in \mathcal{M}(t)}  \| {\bf F}_{\rm RF}(t) {\bf f}_{{\rm BB},j}(t) \|^2 \le P, 
    \nonumber
    \\
    & \hspace{.8cm}
    {\bf f}_{{\rm RF},j}(t) \in {\mathcal F}, \quad j \in \mathcal{M}(t), 
    \label{opt:oneshot:con}
\end{align}
where $w_i(t)=1/{R_i(t-1)}$  is the scheduling weight for user $i$ and we set $R_i(0)=1$ for $i \in \{1,2,...,I\}$. Note that $r_i(t)$ in \eqref{opt:oneshot:obj} is a function of the variables as shown in \eqref{eq:rate}.


\subsection{Challenges for PF Maximization in Hybrid Structure}
\label{ssec:challenges}

The one-shot optimization problem in \eqref{opt:oneshot:obj}-\eqref{opt:oneshot:con} is still challenging to solve for two reasons.
First, the optimization problem is a mixed integer program (MIP), where ${\mathcal{M}}(t)$ and ${\bf F}_{\rm RF}(t)$ exist in the discrete space while ${\bf F}_{\rm BB}(t)$ resides in the continuous space.
The total number of possible combinations for the discrete solution variables is $ \sum_{i=1}^{N_{\max}} \big( {I \choose i}  \times \vert \mathcal{F} \vert^i \big)$.
For a large number of users ($I$) and/or a large codebook ($\vert \mathcal{F} \vert$), the discrete space becomes extremely large.
Second, given the SINR and data rate formulas in \eqref{eq:SINR} and \eqref{eq:rate}, all the channel vectors $\{{\bf h}_i(t)\}_{i=1}^I$ need to be known at the BS to solve the problem. The acquisition of all the channel information requires all of the users to (i) estimate the channels $\{{\bf h}_i(t)\}_{i=1}^I$ and (ii)
feed back the estimated channels to the BS at each time $t$, which 
incurs a large time overhead.


\section{Tailored Two-Timescale Protocol}

To address the challenges, we adopt a two-timescale protocol and tailor it by incorporating an adaptive user selection procedure (Sec.~\ref{ssec:protocol}). We then reformulate the optimization problem based on the modified  protocol (Sec.~\ref{ssec:short_timescale}).




\begin{figure*}[t]
  \includegraphics[width=.8\linewidth]{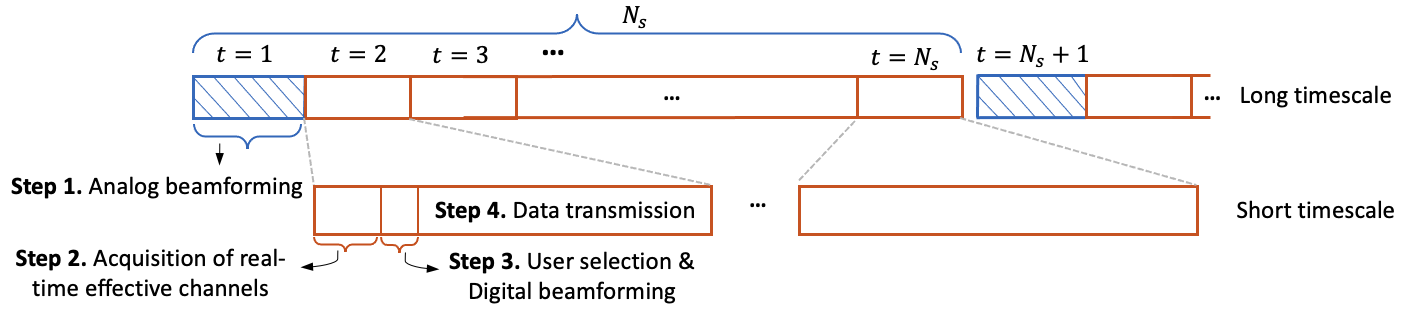}
  \centering
  \caption{Time frame structure of the two-timescale protocol for adaptive user selection in the mmWave hybrid beamforming system.
  }
  \label{fig:protocol_timeline}
\end{figure*}

\subsection{Tailored Two-Timescale Protocol}
\label{ssec:protocol}

In mmWave, the path angles are varying more slowly  than the path gains~\cite{akdeniz2014millimeter,va2016beam}, which motivates the introduction of the two timescales~\cite{liu2019novel,cai2020two}, where the path angles are assumed to be fixed in a long-time block while the path gains change more quickly.
In mmWave channels, there are a few
clusters each with a specific path angle, and thus it is reasonable to utilize a \textit{directional} analog beamformer, such that the maximum gain of the analog beamformer is in the same direction as the dominant path angle\cite{akdeniz2014millimeter}.
This implies that the analog beamforming mostly relies on the slowly-varying path angles in mmWave. On the other hand, the digital beamforming is conducted to adapt to fast-varying path gains.
In this work, we consider the user selection procedure in addition to the analog/digital beamforming.

Therefore, in the tailored two-timescale protocol,
the analog beamforming is conducted in a long timescale,
while the user selection and digital beamforming are performed on a short timescale, which
%
is depicted in Fig. \ref{fig:protocol_timeline}. The short-time block is indexed by $t \in \{1,2,...\}$ and
the long-time block consists of $N_s$ short-time blocks.
%
%
%
%
The detailed procedure
is as~follows: 

\begin{itemize}[leftmargin=5mm] 
    \item \textbf{Step 1. Analog beamforming.} At the beginning of each long-time block\footnote{More time blocks can be allocated for analog beamforming, if needed.}, i.e., $t=1, N_s+1, ...$,
    the BS broadcasts a pilot symbol to a total of $I$ users by using the analog beamformer ${\bf g}_k$ sequentially over $k=1,2,...,\vert \mathcal{F} \vert$, where ${\bf g}_k$ is the $k$-th analog beamforming vector in the codebook $\mathcal{F}$.
    Each user $i$, $i \in \{1,2,...,I\}$, selects the best beam index $k_i(t) \in \{1,2,...,\vert \mathcal{F} \vert\}$, 
    such that
    \begin{align}
        k_i(t) = \underset{ k=1,...,\vert \mathcal{F} \vert} {\arg\max} \;\; \big\vert {\bf h}_i^H(t) {\bf g}_k \big\vert^2.
        \label{eq:analog_beam_index}
    \end{align}
    Each user $i$ feeds back $k_i(t)$ to the BS.
    Then,
    the BS recovers the best analog  beam for user $i$ as ${\bf f}^\star_{{\rm RF},i} = {\bf g}_{k_i(t)}$ from the codebook $\mathcal{F}$. During the next $N_s-1$ short-time blocks, the BS uses the fixed analog beams $\{{\bf f}^\star_{{\rm RF},i}\}_{i \in \{1,2,...,I\}}$.
    \item \textbf{Step 2. Acquisition of real-time effective channels.} At the beginning of every short-time block (e.g., $t=2,3,...,N_s$), the BS explores the analog beamformer ${\bf f}^\star_{{\rm RF},j}$ over $j = 1, 2, ..., I$.
    Each user $i$ measures the effective channel value $u_{i,j}(t) = {\bf h}_i^H(t) {\bf f}^\star_{{\rm RF},j} \in \mathbb{C}$ for $j \in \{1,2,...,I\}$
    and
    feeds back $\{u_{i,j}(t)\}_{j \in \{1,...,I\}}$ to the BS. Receiving the feedback information from all the users, the BS finally obtains  $\{u_{i,j}(t)\}_{i,j \in \{1,...,I\}}$.
    \item \textbf{Step 3. User selection and digital beamforming.} The BS conducts the user selection and digital beamforming based on the real-time effective channels $\{u_{i,j}(t)\}_{i,j \in \{1,...,I\}}$, 
    the scheduling weights $\{w_i(t)\}_{i \in \{1,...,I\}}$, and the analog beamformers $\{{\bf f}^\star_{{\rm RF},j}\}_{j \in \{1,...,I\}}$.
    This step will be discussed in detail in Sec. \ref{ssec:short_timescale}.
    \item \textbf{Step 4. Data transmission.} 
    Using the designed variables, the BS conducts the data transmission with the data rate $r_i(t)$ in \eqref{eq:rate} during the remainder of the short-time block. The BS updates 
    the scheduling weights as $w_i(t+1) = 1/R_i(t)$, where $R_i(t) = (1-\delta)R_i(t-1) + \delta r_i(t)$ as in \eqref{eq:cum_rate}. We conduct \textbf{Step 2-4} until the end of the long-time block.
\end{itemize}

\subsection{Optimization for  User Selection and Digital Beamforming}
\label{ssec:short_timescale}

Using the modified two-timescale protocol, 
we have reduced the design complexity for analog beamforming by selecting the best analog beam for each user only once in each long-time block. Then, once the user set $\mathcal{M}=\{i_1, ...,i_M\}$ is determined for a short-time block, the analog precoder will be designed~as\footnote{We remove the time index $t$ from this subsection, since the optimization  is conducted with the corresponding variables  for every short-time block.}
\begin{align}
    {\bf F}^\star_{\rm RF}(\mathcal{M}) = [ {\bf f}^\star_{{\rm RF},i_1}, ...,  {\bf f}^\star_{{\rm RF},i_M}].
\end{align}

We then simplify the optimization problem \eqref{opt:oneshot:obj}-\eqref{opt:oneshot:con} as
%
%
%
%
%
%
%
%
%
\begin{align}
    & \underset{{\mathcal{M}}, \;
    {\bf F}_{\rm BB}}{\rm maximize} \quad \sum_{i=1}^I w_i r_i
    \label{opt:simplified:obj}
    \\
    & {\rm s.t.} \quad 
    {\mathcal{M}} \subset \{1,...,I \}, \quad \vert {\mathcal{M}} \vert \le N_{\max}, 
    \nonumber
    \\
    & \hspace{.8cm} 
    \sum_{i \in \mathcal{M}}  \| {\bf F}^\star_{\rm RF}(\mathcal{M}) {\bf f}_{{\rm BB},i} \|^2 \le P,
    \nonumber
    \\
    & \hspace{1mm}
    r_i \negmedspace=\negmedspace 
    \begin{cases}
    \log_2 \bigg(
    1 + \frac{  \big\vert {\bf u}_i^H(\mathcal{M})  {\bf f}_{{\rm BB},i}  \big\vert^2 }
    {\sum\limits_{j \in \mathcal{M}\setminus \{i\} }  \big\vert {\bf u}_i^H(\mathcal{M})  {\bf f}_{{\rm BB},j}  \big\vert^2 +   \sigma_i^2  }
    \bigg), & \negmedspace i \in \mathcal{M} \\
    0, & \negmedspace i \notin \mathcal{M}
    \end{cases}
    \label{opt:simplified:con}
\end{align}
where
the last constraint in \eqref{opt:simplified:con} is the data rate expression from \eqref{eq:SINR} and \eqref{eq:rate}. In \eqref{opt:simplified:con}, ${\bf u}_i^H(\mathcal{M}) = {\bf h}_i^H {\bf F}^\star_{{\rm RF}}(\mathcal{M}) = [u_{i,i_1}, u_{i,i_2}, ..., u_{i,i_M}]$ is the effective channel vector from the BS to user $i$, when ${\bf F}^\star_{\rm RF}(\mathcal{M})$ is used for the analog precoder.


\subsection{Zero-Forcing (ZF) Digital Beamforming}

The optimization problem \eqref{opt:simplified:obj}-\eqref{opt:simplified:con} is a joint user selection and digital beamforming problem.
Once
$\mathcal{M}=\{i_1,...,i_M\}$ is determined, ${\bf F}_{\rm BB}$ can be designed by solving  \eqref{opt:simplified:obj}-\eqref{opt:simplified:con}
 for ${\bf F}_{\rm BB}$.
To design ${\bf F}_{\rm BB}$,
we adopt a zero-forcing (ZF) beamforming, which is shown 
to perform well in  mmWave channels due to a low number of clusters~\cite{alkhateeb2015limited,akdeniz2014millimeter}.

Now, we will briefly describe the ZF beamforming to design ${\bf F}_{\rm BB}$ when $\mathcal{M}=\{i_1,...,i_M\}$ is given.
We first define 
the effective channel matrix associated with the selected users as
\begin{align}
    {\bf G}(\mathcal{M}) = 
    \begin{bmatrix}
        {u}_{i_1,i_1} & ... &  {u}_{i_1,i_M}\\
        \vdots & \ddots & \vdots \\
        {u}_{i_M,i_1} & ... & {u}_{i_M,i_M} 
    \end{bmatrix}
    \in \mathbb{C}^{ M \times M}.
    \label{eq:G}
\end{align}
For ZF beamforming, we determine ${\bf F}_{\rm BB}$ such that
${\bf G}(\mathcal{M}) {\bf F}_{\rm BB} = {\bf I}$, i.e., 
\begin{align}
    {\bf F}_{\rm BB}(\mathcal{M}) 
    = 
    ({\bf G}(\mathcal{M}))^H ({\bf G}(\mathcal{M})({\bf G}(\mathcal{M}))^H)^{-1}.
    \label{eq:FBB:inv}
\end{align}

We note that ${\bf F}_{\rm BB}(\mathcal{M}) = [ {\bf f}_{{\rm BB},i_1}, ..., {\bf f}_{{\rm BB},i_M}]$. 
We consider an equal power allocation for the $M$ data streams as in \cite{alkhateeb2015limited}, i.e., $\| {\bf F}^\star_{\rm RF}(\mathcal{M})  {\bf f}_{{\rm BB},i} \|^2 = \frac{P}{M}$, $i \in \mathcal{M}$.
To this end, we normalize each digital beamforming vector in ${\bf F}_{\rm BB}(\mathcal{M})$ and obtain 
\begin{align}
    {\bf f}^\star_{{\rm BB},i} =  \sqrt{\frac{P}{M}} \frac{ {\bf f}_{{\rm BB},i}} {\| {\bf F}^\star_{\rm RF}(\mathcal{M})  {\bf f}_{{\rm BB},i} \|_2}, \quad i \in \mathcal{M}.
    \label{eq:fBB_star}
\end{align}
By collecting the obtained beamformers, we finally construct the digital precoder as
${\bf F}^\star_{\rm BB}(\mathcal{M}) = [{\bf f}^\star_{{\rm BB},i_1}, ..., {\bf f}^\star_{{\rm BB},i_M}]$.
With this,
the weighted sum-rate in \eqref{opt:simplified:obj} is
given by 
\begin{equation}
    \hspace{-2mm} \resizebox{0.92\hsize}{!}{$
    Q(\mathcal{M}) \negmedspace = \negmedspace \sum\limits_{i \in \mathcal{M}} w_i \log_2 \bigg(
    1 + \frac{  \big\vert {\bf u}_i^H(\mathcal{M})  {\bf f}^\star_{{\rm BB},i}  \big\vert^2 }
    {\sum\limits_{j \in \mathcal{M}\setminus \{i\}}  \big\vert {\bf u}_i^H(\mathcal{M})  {\bf f}^\star_{{\rm BB},j}  \big\vert^2 +   \sigma_i^2 }
    \bigg).$}
     \hspace{-2mm}
    \label{eq:Q_M}
\end{equation}

We aim to select the users in a way that maximizes the weighted sum-rate in \eqref{eq:Q_M}.
A naive approach is to conduct the exhaustive search by calculating the ZF digital precoder for every possible $\mathcal{M}$ and select the
best case
that yields the largest $Q(\mathcal{M})$. 
However, an exhaustive search would incur a large computational overhead once the number of possible $\mathcal{M}$ is large.
Instead, we can exploit simpler solution methods, such as a greedy algorithm and a top-$k$ algorithm, which we will discuss in the next section.

\section{Combinatorial Solvers for User Selection}
\label{sec:comb}

In this section, we  
present two different combinatorial solvers, the greedy algorithm in Sec.~\ref{ssec:greedy} and the top-$k$ algorithm in Sec.~\ref{ssec:topk},
to
solve the optimization problem \eqref{opt:simplified:obj}-\eqref{opt:simplified:con}.

 \subsection{Greedy Algorithm}
 \label{ssec:greedy}
The underlying idea of the greedy algorithm is to start from an empty set and add a user one by one to the set until the sum-rate performance is not further improved.
Using $Q(\mathcal{M})$ in \eqref{eq:Q_M},
we can represent the greedy algorithm in Algorithm \ref{al:greedy}.
Starting from an empty set $\mathcal{M} = \{ \}$, we first find user $i^\star$ 
that yields the highest weighted rate, i.e., $i^\star  = \arg\max_{i} Q( \{ i \} )$, and add $i^\star$ to the set $\mathcal{M}$. 
In the next round, we find the next user $i^\star$ 
that yields the highest weighted sum-rate over the incorporated user set, i.e., $i^\star = \arg\max_i Q( \mathcal{M} \cup \{ i \} )$. 
If adding user $i^\star$ to the set yields a better performance, i.e.,
$Q(\mathcal{M} \cup \{i^\star\})  >  Q(\mathcal{M})$, we add $i^\star$  to $\mathcal{M}$. 
We keep doing this process until 
$|\mathcal{M}|$ reaches to $N_{\rm max}$.
If $Q(\mathcal{M} \cup \{i^\star\})  \le  Q(\mathcal{M})$, we stop the~algorithm.

 \begin{algorithm}[h!]
 \caption{Greedy algorithm for user selection}
 \label{al:greedy}
 \begin{algorithmic}[1]
 \small
\State \textbf{Input.}  
 $I$,   $N_{\max}$,   
 $\{ w_i \}_{i \in  \{ 1,...,I \}}$, $\{{\bf f}^\star_{{\rm RF},i}\}_{i \in  \{ 1,...,I \}}$, and  $\{{u}_{i,j}\}_{i,j \in \{ 1,...,I \}}$.
 \State \textbf{Output.} The set of the selected users, $\mathcal{M} \subset \{1,...,I\} $.
  \State Initialize $\mathcal{M} = \{ \}$, $\overline{\mathcal{M}} = \{1,...,I\}$, and
  $Q(\mathcal{M}) = 0$.
  \For {$n = 1, ..., N_{\max}$}
    \State \multiline{ 
    Determine the best user $i^\star$ in the set $\overline{\mathcal{M}}$ so that the weighted sum-rate is maximized, i.e.,
        $i^\star = \underset{i \in \overline{\mathcal{M}}}{\arg\max} \; Q(\mathcal{M} \cup \{i\})$.}
    \label{greedy:line:comp}
    \If{$Q(\mathcal{M} \cup \{i^\star\}) \le Q(\mathcal{M})$}
    \State \textbf{break}
    \EndIf
    \State Update the sets as $ \mathcal{M} \leftarrow \mathcal{M} \cup \{i^\star\} $ and $ \overline{\mathcal{M}} \leftarrow \overline{\mathcal{M}} \setminus \{i^\star\} $.
     \EndFor
 \end{algorithmic}
 \end{algorithm}

 \subsection{Top-$k$ algorithm}
 \label{ssec:topk}
The greedy algorithm is still computationally expensive due to calculating the ZF precoder for many different user sets.
To further reduce the computation overhead, we exploit the top-$k$ algorithm for user selection. 
We first calculate the data rate of each user based on the effective channels $\{{u}_{i,i}\}_{i \in \{ 1,...,I \}}$ and the scheduling weights $\{ w_i \}_{i \in  \{ 1,...,I \}}$ without considering the interference among the users as
$ \bar r_i = w_i \log_2 \big( 1 + P \vert  u_{i,i} \vert^2/\sigma_i^2 \big)$, $i \in \{1,...,I\}$.
Then, we rank the data rates in large order as $\bar r_{i_1} \ge \bar r_{i_2} \ge ... \ge \bar r_{i_I}$. We finally select the $k$ best users and construct 
$\mathcal{M} = \{i_1, ..., i_k \}$.
The top-$k$ algorithm is computationally efficient since it does not include the process for the calculation of the ZF digital precoder for user selection.
Over the selected users $\mathcal{M} = \{i_1, ..., i_k \}$, the BS conducts the ZF digital precoding. 
In summary, once $k$ is given, we can implement the top-$k$ algorithm to select the $k$ users.
%
Furthermore, we can develop an advanced version of the top-$k$ algorithm by running the top-$k$ algorithm for different $k$, $k=1,...,N_{\max}$, and choosing the best $k$. We call this the \textit{adaptive} top-$k$ algorithm.

\section{Learning-Based User Selection}

We note that the greedy algorithm does not operate fast enough to optimize for the variables within the short time period in mmWave systems, i.e., within a few ${\rm msecs}$.
On the other hand, the top-$k$ algorithm 
does not yield a good PF performance.
Machine learning (ML) has been exploited in many domains of literature to provide a fast inference after the machine is trained, and to yield a good performance by capturing implicit features in the observed data.
Motivated by this, we propose to exploit deep neural networks (DNNs). 

We consider a fully connected neural network with two hidden layers, which have $L_1$ and $L_2$ neurons, respectively.
Generally, the input of the DNN consists of $I^2$ effective channels $\{u_{i,j}\}_{i,j \in \{1,...,I\}}$,  $I$ scheduling weights $\{w_i\}_{i \in \{1,...,I\}}$, and $I$ analog beam indices $\{ k_i \}_{i \in \{1,...,I\}}$.\footnote{We study different DNN architectures, e.g., different number of layers and neurons and different input combinations, in Sec.~\ref{ssec:study_NN_architecture}.}
For the output, we define  the user selection vector as
${\bf a} = [a_1, ..., a_I] \subset \{ 0,1 \}^{I}$. If $a_i =1$, user $i$ is selected. 
Then, we can construct $\mathcal{M} = \{ i: \; a_i=1, \; i =1,...,I \}$.
We use a sigmoid activation fuction at each neuron in the output layer.

For inference, we 
consider two additional steps after the output layer.
First, we round each output of the sigmoid function to $0$ or $1$.
This binary value then corresponds to the decision of the user selection.
Second, if the number of selected users inferred from the DNN, $\vert {\mathcal{M}} \vert$, is larger than $N_{\rm max}$, 
we remove extra users with lower data rates one by one through the reverse process of the top-$k$ algorithm (described in Sec.~\ref{sec:comb}), 
until $\vert {\mathcal{M}} \vert=N_{\rm max}$. 
This enables us to satisfy the constraint on the number of selected users.
For training, we consider $N_e$ episodes, where each episode contains  $T$ timesteps.
Each episode has different realizations of the initial channels and user distributions/moving directions.
To obtain the input/output as training data, we run the greedy algorithm (Algorithm \ref{al:greedy}) for every short-time block.
Specifically, as an input, we obtain (i) the analog beam indices (from \textbf{Step 1}), (ii) the real-time effective channels  (from \textbf{Step 2}), and (iii) the scheduling weights (from \textbf{Step 4}).
For the output, we obtain the user selection set $\mathcal{M}$ by running the greedy algorithm,
and constructing the target vector ${\bf a} = [a_1, ..., a_I]$ by setting $a_i=1$ for $i \in \mathcal{M}$ and $a_i=0$ for $i \notin \mathcal{M}$.
The total number of input-output pairs used for training is $N_e T$.
\section{Numerical Results}

We first describe the simulation setup in Sec.~\ref{ssec:setup}, and then
compare the ML-based algorithm with other baselines under different metrics
in Sec.~\ref{ssec:performance}.
Lastly, we study the impact of employing different DNN architectures in Sec.~\ref{ssec:study_NN_architecture}.

\begin{figure}[t]
%
\centering
\begin{subfigure}{\linewidth}
  \centering
  \includegraphics[width=.9\linewidth]{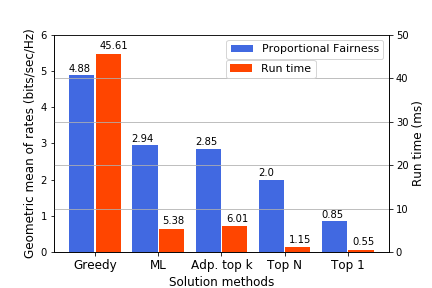}
  \caption{ Proportional fairness and run time.
  }
  \label{fig:metric1}
\end{subfigure}
\begin{subfigure}{\linewidth}
  \centering
  \includegraphics[width=.9\linewidth]{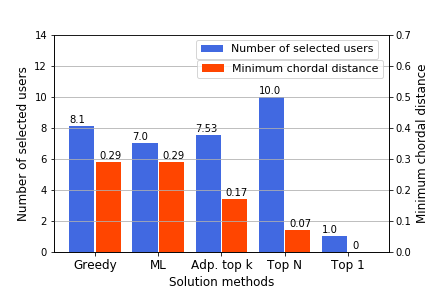}
  \caption{ Number of selected users and  channel distance.
  }
  \label{fig:metric2}
\end{subfigure}
\begin{subfigure}{\linewidth}
  \centering
  \includegraphics[width=.9\linewidth]{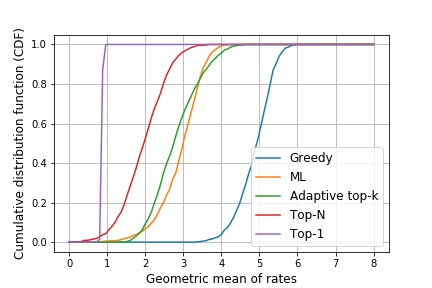}
  \caption{CDF of geometric mean of rates.
  }
  \label{fig:CDF}
\end{subfigure}
\caption{(\subref{fig:metric1}), (\subref{fig:metric2}) Comparison among different methods under various metrics. (\subref{fig:CDF}) CDF plot of the PF metric obtained from different methods.
}
\label{fig:comparison}
\end{figure}

\subsection{Simulation Setup}
\label{ssec:setup}

We set $I=20$, $N_{\max}= 10$, $P=2$ Watt, $f_c=28$ GHz, $\delta = 0.1$, and $\sigma_i^2 = 10^{-15}$, $i \in \{1,...,I\}$.
We consider a uniform planar array (UPA) of the antennas at the BS, where $N_{\rm BS}=16$ with $N_{{\rm BS},x}=8$ in horizontal and $N_{{\rm BS},y}=2$ in vertical.
The BS is located at $7$m height~\cite{azar201328}, while $I$ users are randomly distributed around the BS with radius $100$m. The boresight of the BS antenna array is tilted downward at $10^o$. 
The BS scans $(-180^o,180^o)$ horizontally and $(-30^o,30^o)$ vertically.
The codebook $\mathcal{F}$
is derived from a $32\times 8$ grid of beams that is equally spaced in both the horizontal and vertical directions. 
We assume the user velocity $v=4$ km/h, and
adopt the mmWave channel model given in~\cite{akdeniz2014millimeter}. We use Pytorch to implement the ML-based algorithm.

For the DNN, unless otherwise stated, we consider $L_1=500$ and $L_2=200$ neurons, where
the input consist of the magnitude of the effective channels, $\{\vert u_{i,j} \vert\}_{i,j \in \{1,...,I\}}$ and the weights $\{w_i\}_{i \in \{1,...,I\}}$. 
The size of the input is then $I^2 + I$.
For training, we set $N_e = 12000$ episodes and $T=120$ timesteps.
We consider $N_s=40$.
The number of epochs is $300$, where 
the epoch means one complete pass of the entire training dataset with size $N_eT$. We adopt the binary cross entropy loss and the Adam optimizer for training.
We normalize the input of the training data separately for the weights and the effective channels because their orders of magnitudes are different.
The accuracy is calculated element-by-element, i.e., 
    ${\rm acc} = 1- {\sum_{i=1}^I \vert {a}_{i,{\rm predict}} - {a}_{i,{\rm target}} \vert}/{I}$,
where ${a}_{i,{\rm predict}} \in \{ 0,1\}$  denotes the predicted value at the $i$th neuron in the output layer, while  ${a}_{i,{\rm target}} \in \{ 0,1\}$ denotes the target value.
For testing, we have $N_e = 1000$ and $T=120$.


\subsection{Performance Comparison under Different Metrics}
\label{ssec:performance}


We compare different user selection algorithms: the greedy algorithm, ML-based algorithm, and variations of the top-$k$ algorithm, where the top-$1$ algorithm selects the best single user, while the top-$N$ algorithm selects $N_{\rm max}=10$ users. For the ML-based algorithm, the accuracy is 71\%. Although the accuracy value is not high, the PF obtained from the trained DNN is comparable to other methods with a reasonable running time, which will be discussed in the following.

In Fig.~\ref{fig:comparison}(\subref{fig:metric1}), we consider two metrics, the proportional fairness (PF) 
and the run time. 
We note that the geometric mean of rates is
another representation of the PF metric, 
defined by
$(\prod_{i=1}^I R_i(T))^{1/I}$.
As expected, the greedy algorithm yields a good PF performance, while it requires a high computation time, which may not be realistic to operate in every short-time block. On the other hand, the  top-$N$ and top-$1$ algorithm yields a short run time, while its obtained PF performance is not satisfactory.
The adaptive top-$k$ algorithm provides a decent PF performance with higher run time than the top-$N$ and top-$1$ algorithm. 
The ML-based algorithm provides a better performance than the adaptive top-$k$ algorithm with higher PF and lower run time. 
This shows that the ML-based algorithm gives an efficient trade-off between the PF and the run time.


In Fig.~\ref{fig:comparison}(\subref{fig:metric2}), we consider two other metrics, the number of selected users 
and the minimum chordal distance.
The greedy, ML-based, and adaptive top-$k$ algorithm select around 7 to 8 users on average.
It is interesting that the ML-based algorithm yields a high value of the minimum chordal distance (similar to the greedy algorithm), implying that the channel vectors of the selected users are well separated from each other.
This means that the ML-based algorithm learns to mitigate the channel interference of the users.
Fig.~\ref{fig:comparison}(\subref{fig:CDF}) shows the cumulative distribution function (CDF) of the PF. 
We note that the greedy algorithm 
requires a long run time that could not be executed in real-time.
The ML-based method yields the second best PF performance with a reasonable run time.



\subsection{Study on Different DNN Architectures}
\label{ssec:study_NN_architecture}

\begin{table}
\caption{Different DNN input types (The unit of run time is msec).}
\centering
\scalebox{0.75}{
\begin{tabular}{
|c||c|c|c|c|c|c|c|  }
 \hline
 & W & W+C(D) & W+C(D)+B & C(W) & W+C(W) & W+C(W)+B & W+C(R/I) \\
 \hline
 PF  &1.36 &1.77 &1.76 & 2.54 & {\bf 2.94} & 2.92 & 1.30\\
 \hline
 Run time  &5.22 &5.3 &5.31 & 5.36 & 5.38 & 5.42 & 5.48\\
 \hline
\end{tabular}}
\label{table:input}
\end{table}

\begin{table}
\caption{Different number of layers and neurons for the DNN.}
\centering
\scalebox{0.85}{
\begin{tabular}{
|c||c|c|c|c|c|  }
 \hline
 & $50 \negmedspace \times \negmedspace 20$ & $100 \negmedspace \times \negmedspace 50$ & $200 \negmedspace \times \negmedspace 100$ & $500 \negmedspace \times \negmedspace 200$ & $500 \negmedspace \times \negmedspace 200 \negmedspace \times \negmedspace 100$ \\
 \hline
 PF  & 2.63 & 2.81 & 2.81 & ${\bf 2.94}$ & 2.84 \\
 \hline
 Run time  & 4.93 & 5.22 & 5.24 & 5.38 & 5.89\\
 \hline
\end{tabular}}
\label{table:layer}
\end{table}

We compare the performances obtained from the ML-based algorithm trained under different DNN architectures. 
First, we study different combinations of
the DNN input.
In Table~\ref{table:input}, C, W, and B denotes effective channels, scheduling weights, and beam indices, respectively.
Also, C(R/I) denotes the use of the real and imaginary values for the effective channels.
The C(D) denotes the use of only the diagonal magnitude values, i.e., $\{\vert u_{i,i} \vert \}_{i \in \{1,...,I\}}$, 
while C(W) denotes the whole magnitude values, $\{\vert u_{i,j} \vert \}_{i,j \in \{1,...,I\}}$.
%
Table \ref{table:input} shows that using the weights and whole effective channels (without the analog beam indices) yields the best PF performance. The run time is almost similar for all the cases because the run time is 
not dominated by the DNN inference, rather by 
the design of the ZF digital precoder and the data conversion from GPU to CPU.

Table~\ref{table:layer} shows the study of different number of layers and neurons.
Using many layers and neurons may lead to an over-fitting during training due to its high model complexity, while a small DNN with lower number of layers and neurons would not capture the implicit features properly from the input data. 
It is shown that using two hidden layers with 500 and 200 neurons in each layer, respectively, yields the best PF.
\section{Conclusion}

We
 tailored the signal model for mmWave hybrid beamforming systems by incorporating the varying number of users.
We then formulated the PF maximization problem for dynamic user selection.
To address the challenges in designing the variables,
we adopted an efficient two-timescale protocol
and 
incorporated
the user selection procedure into the protocol.
We first exploited the greedy algorithm and top-$k$ algorithm for adaptive user selection.
Considering the trade-off between the PF performance and the computational complexity, we proposed a ML-based user selection algorithm. 
Through simulations, we demonstrated that the ML-based algorithm gives an efficient trade-off between the PF performance and the run time, compared to the other solution methods.
In future work, it will be interesting to further improve the ML-based
algorithm to significantly outperform the baselines in terms of
various metrics discussed in this work.



\bibliographystyle{IEEEtran}
\balance
\bibliography{ref}

\end{document}